\documentstyle[epsf]{mn}
\newcommand{\HI}{\hbox{H\,{\sevensize I}}}
\newcommand{\DDO}[1]{DDO~#1}
\newcommand{\NGC}[1]{NGC~#1}
\newcommand{\Msun}{M_\odot}
\newcommand{\Zsun}{Z_\odot}
\begin{document}
\title[\DDO{154}]{On \DDO{154} and Cold Dark Matter halo profiles}
\author[S. Gelato and J. Sommer-Larsen]{Sergio~Gelato and Jesper
Sommer-Larsen\\
Theoretical Astrophysics Center, Juliane Maries Vej 30, 
2100 Copenhagen \O{}, Denmark}
\maketitle
\begin{abstract}
We investigate the claim by Burkert and Silk \shortcite{BS97} that the
observed rotation curve of the dwarf irregular galaxy \DDO{154}
cannot be reconciled with the universal CDM halo profile of Navarro,
Frenk \& White \shortcite{NFW96,NFW97} even when allowance is made for
the effect of violent gas outflow events on the structure of
the galaxy.
By means of $N$-body simulations we show that under certain conditions
it is possible to obtain a reasonable fit to the observed rotation
curve without invoking Burkert \& Silk's proposed spheroidal MACHO component.
We are able to best reproduce the observed decline in the rotation
curve by postulating additional hidden disc mass, in an amount 
that is compatible with disc stability requirements.
In the process we improve upon the results of Navarro, Eke
\& Frenk \shortcite{NEF96} on the formation of halo cores by mass
loss by using actual haloes from Cold Dark Matter simulations
instead of Hernquist \shortcite{H90} distributions.
\end{abstract}

\section{Introduction}

A rather generic feature of $N$-body models of collisionless
gravitational collapse is the presence of a central density cusp
in the objects that result.
While such cusps are observed in many elliptical galaxies,
and while in large high-surface-brightness spirals the dominant
disc component makes them difficult to discern unambiguously,
they are conspicuously absent from the profiles one infers
from the observed rotation curves of dwarf disc galaxies
\cite{M94,FP94}.
Instead, the dynamics of these objects imply a dark matter distribution
characterized by an extended core of nearly uniform density.
The nearby galaxy \DDO{154} is a prime example:
Carignan \& Freeman \shortcite{CF88} find a well-constrained fit of
the dark matter component by an isothermal sphere with a core radius
$r_c=3.0$~kpc.

Navarro, Eke and Frenk \shortcite[hereafter NEF]{NEF96} have shown how
the central cusp could have been erased some time after the halo was
assembled, by the violent relaxation that follows a sudden large, 
presumably starburst-driven, mass ejection event.
In their numerical $N$-body experiments they added a slowly
growing disc potential to the halo, then removed it abruptly.
The slow growth of the disc causes the inner parts of the halo to
contract adiabatically, and increases the velocity dispersion there.
The subsequent sudden mass loss reduces the binding energy in the
central region, leading on a dynamical time scale to the establishment
of a new equilibrium with a core in the density profile.

NEF assumed a Hernquist \shortcite{H90} distribution function
for the halo at the beginning of the disc growth phase.
This distribution has a weak ($\rho \propto r^{-1}$) central cusp
similar to that reported by Dubinski \& Carlberg \shortcite{DC91}
and consistent with the ``universal halo profile'' of Navarro, Frenk
\& White \shortcite[hereafter collectively NFW]{NFW96,NFW97}.
The density profile of the Hernquist distribution falls off more
quickly (as $r^{-4}$ rather than $r^{-3}$) at large radii, however;
this difference has a bearing on the behaviour of the rotation curve
at large galactocentric distances (which was not the focus of NEF's
study) and also affects the velocity structure within the halo.
Perhaps more importantly, the universality of $r^{-1}$ cusps in CDM
haloes is somewhat controversial: various groups \cite{FM97,M.97}
report steeper slopes in higher-resolution simulations with finer mass
granularity.
These stronger cusps are naturally more difficult to obliterate (see 
our results below).

NFW claim that in a wide range of Cold Dark Matter (CDM) cosmogonies
the radial profiles of virialised haloes are well described by
\begin{equation}
\rho(r) = {\delta_c \rho_c r_s^3 \over r (r_s+r)^2}
\label{q:nfw1}
\end{equation}
where $\rho_c$ is the mean density of the Universe, $\delta_c$ a
characteristic density enhancement, and $r_s$ a characteristic radius
around which the profile changes from a logarithmic slope of $-1$ to
one of $-3$.
For a halo of a given mass $M_{200}$ within the virial radius
$r_{200}$ (which, following common practice, we define as the radius
of a sphere
within which the mean density is $200\rho_c$), $\delta_c$ is related
to the  ``concentration parameter'' $c \equiv r_{200}/r_s$ by
\begin{equation}
\delta_c = {200\over 3} {c^3 \over [\ln(1+c)-c/(1+c)]} .
\label{q:nfw2}
\end{equation}
Since in a CDM cosmogony haloes of a given mass are preferentially
assembled in a fairly narrow range of redshifts (with smaller haloes
forming first), there is a correlation between $\delta_c$ (or~$c$)
and~$M_{200}$: less massive haloes are denser and more centrally
concentrated.

In the case of \DDO{154}, a very isolated dwarf disc galaxy whose \HI{}
rotation curve has been measured to a considerable distance from its centre
\cite{CF88,H.93}, the ``universal profile'' of NFW cannot
simultaneously account for the observations at small and at large radii.
If the parameters ($\delta_c$, $r_s$) are chosen to
fit the observations at small radii (after allowance has been made for
the formation of a uniform-density core by mass outflow), the NFW 
profile predicts that the
rotation curve should continue to rise well beyond the radius at which
it is observed to peak.
If on the other hand the parameters are adjusted to
fit the declining part of the rotation curve, the NFW profile predicts
an unreasonably large mass excess in the central regions.
Burkert \& Silk \shortcite[hereafter BS]{BS97} have invoked this
difficulty as an
argument for the presence of an additional dark component formed
through dissipative processes, analogous to the putative MACHO
component of our own Galaxy.
In their view, this new component should be spheroidal rather than
disc-like since an overly massive disc would be unstable to
fragmentation and expected to form large quantities of stars, contrary
to what is observed in this galaxy.
They still assume that before this dissipational collapse process the
halo was well described by the NFW profile.

\DDO{154} is not the only dwarf galaxy for which the observational data
are at variance with the cuspy profiles predicted by most $N$-body
simulations.
Burkert~\shortcite{B95} and Kravtsov et~al~\shortcite{K.98} have
argued that the observational data can be described by a different
one-parameter family of profiles with at most a very weak cusp.
Of the galaxies examined in these two works, however, only \DDO{154} and
\NGC{2915} have the very extended rotation curves required for a
really stringent test of secular evolution models for the halo density
profile.
\NGC{2915} \cite{M.96} is in many ways a peculiar object, making it
much more difficult to model.
We therefore follow the example of Burkert \& Silk~\shortcite{BS97}
and concentrate on \DDO{154} alone.

In this paper we explore the possibility of achieving a reasonable
fit to the observed rotation curve of \DDO{154} without invoking such
a dissipative spheroidal component.
Instead, we experiment with different mass ejection fractions and disc
density distributions.
We improve on the work of NEF by using halo distribution functions
drawn from ab initio cosmological simulations: the Hernquist
\shortcite{H90} distribution's unrealistically fast drop-off at
large radii might yield a spuriously better fit to the falling
part of the rotation curve.
Also, NEF's preferred fits for \DDO{154} involve characteristic scale
lengths ($a_h$ in their notation, corresponding roughly to $r_s$ in
this paper) an order of magnitude larger than expected for the halo
of \DDO{154} if one assumes a total mass of order~$10^{10}\Msun$ or less,
commensurate with the observed radial extent and rotation velocity of
this galaxy. (A larger total mass would likely only 
exacerbate the difficulty in reproducing the observed decline in the
rotation curve.)
Alternatively, NEF's disc scale lengths are unreasonably
small (1--4\%) by comparison to the halo scale length. In the present work,
by contrast, the characteristic halo radius $r_s$ is determined by our
ab initio cosmological simulations, and is in good agreement
with the results of NFW; our disc scale lengths are also based on
observations.

We study both models in which the present-day disc mass is 
as implied by optical and \HI{} observations of \DDO{154} \cite{CF88}
and models with additional baryonic mass.
In the latter case, the additional mass is assumed to be hidden in some
hitherto undetected form and distributed like the observed \HI{}.
(The mean metallicity of \DDO{154}, as measured by the O/H abundance
ratio, is only $0.05 \Zsun$ \cite{vZ.97}, and
at such a low metallicity there are both theoretical \cite{MB88} and
observational \cite{VH95} indications that the CO-to-H$_2$ conversion
ratio should be very large, implying that significant amounts of H$_2$
could be present without conspicuous associated CO emission.)
We examine the argument that a more massive disc would be
Toomre-unstable and form stars at a higher rate than observed, and
find that we can accommodate a disc up to three times more massive than
reported from \HI{} observations without having to assume an
implausibly strong stabilizing pressure within the disc.

Section~\ref{s:simulations} describes in detail the parameters of our
$N$-body simulations. 
Results are presented in section~\ref{s:results}, 
and conclusions drawn in section~\ref{s:conclusions}.

\section{The simulations}
\label{s:simulations}

\subsection{Initial conditions}

We selected initial conditions out of a realization of a standard CDM
power spectrum ($\Omega=1$, $\Lambda=0$,
$h\equiv H_0/(100\,\rm{km}\,\rm{s}^{-1}\,\rm{Mpc}^{-1})=0.5$) normalized to
$\sigma_8=0.6$. 
This realization was computed on a $128^3$ grid in a periodic box of only
$4 h^{-1}$~Mpc on a side; this would be too small for studies of
clustering, or even of larger single galaxies, but turned out to be
adequate for our purpose of obtaining haloes matching the profile of
equation~\ref{q:nfw1}.

The box was evolved to the present epoch with a standard particle-mesh
code, and haloes
within the desired range of virial masses were identified in the final
state.
The corresponding regions within the initial conditions were then
resampled using a larger number of particles.
We used $4^3$ particles per grid zone in the high-resolution region
centred on the halo of interest (except for one run with $5^3$
particles per zone, to verify that collisional relaxation was
negligible). The tidal field was sampled out to a radius of 8~Mpc
(using periodic replicas of the original box as needed)
with progressively heavier particles at larger distances.
Each particle in the high-resolution region thus had a mass
of~$2.65\times 10^{5} \Msun$.
The high-resolution region was chosen to include those mass elements
which in the final state ($z=0$) of the particle-mesh simulation belonged to
the chosen halo (within a local overdensity contour $\delta > 55$,
corresponding to a mean enclosed overdensity of about~180 in a
spherical collapse model).
For our haloes, this region typically contains about~$10^{10} \Msun$,
or 40000~particles.

\subsection{Initial collapse phase}

The resampled initial conditions were evolved for 3.0~Gyr
starting from a redshift~$z_i=49$ using a binary tree $N$-body code.
We used a modified version of the code described in Navarro
\& White \shortcite{NW93}, in which the time step selection criterion
was made tighter by comparing the third-order Runge-Kutta-Fehlberg
terms to the first-order position and velocity increments
$|\delta\bmath{x}|$ and $|\delta\bmath{v}|$ rather than to the
zeroth-order, non-Galilean-invariant positions and velocities
$|\bmath{x}|$ and~$|\bmath{v}|$.
For this work we did not make use of the code's built-in Smoothed
Particle Hydrodynamics capability, but operated it solely as
a gravitational $N$-body integrator.
Each particle was given a gravitational softening length proportional
to the cube root of its mass, with high-resolution particles having
a softening length of 0.5~kpc, large enough to avoid two-body
relaxation effects yet still comfortably smaller than the length
scales (such as the 3.0~kpc dark halo core radius) we wish to resolve.

We applied this procedure to 8 of the approximately 20 haloes found within
our preferred mass range, then selected three haloes with
promising rotation curves for further study.
(Although we subjectively deem these rotation curves ``promising'',
they do not match the observational data particularly well. This fact
is illustrated in figure~\ref{f:vc3} and quantified in
table~\ref{t:runs}, both of which are further described below.)
Some scatter in halo properties is expected even in principle, and
for want of observational data we do not know where the properties of
\DDO{154} lie with respect to the ensemble average for objects of that
class.
In looking for a match to the observed rotation curve we can therefore
allow ourselves some reasonable number of tries with different halo
realizations.

\subsection{Disc growth}

At $t=3.0$~Gyr (except for our first halo, number~99, which was
evolved to $t=3.7$~Gyr; both times were chosen somewhat arbitrarily)
we cut a sphere of radius 50~kpc ($\ga 2.5 r_{200}$) around the centre
of the chosen halo,
and applied a Galilean transformation to bring the centre of mass to
rest at the origin of coordinates and to align the total angular
momentum along the $z$~axis.
This was done for practical convenience, as we would otherwise have
been forced to incorporate a feedback mechanism to change the position
and orientation of the growing disc in response to long-range tidal
forces.
As it is, we were able to reuse the existing implementation of NEF,
in which the disc is symmetric about the $z$~axis and the $(x, y)$~plane.
We verified that the density maximum in our simulated haloes does in
fact remain close to the origin (to within our target accuracy of
$\sim 0.5$~kpc) during the course of our subsequent runs.

We impose a slowly growing external potential corresponding
to a thin axisymmetric disc with surface density $\Sigma(R)$, where
$R$ is the radial coordinate within the disc.
The corresponding gravitational potential and acceleration were
computed on a grid by a Hankel transform technique \cite[section 2.6.3]{BT87}.
A Fast Hankel Transform algorithm \cite{A82} allows us to
specify essentially arbitrary surface density profiles.
In the disc growth phase, however, we only use
exponential density distributions $\Sigma(R) \propto e^{-R/R_d}$ for
various values of the scale length~$R_d$.

The disc growth phase lasts $1.5$~Gyr, during which the disc gains
mass at a constant rate at the expense of the particles in the
simulation. We decrease the mass of all the dark matter particles by
the same proportion, in such a way that the sum of the masses of halo and
disc is conserved. This effectively means that the baryonic disc
is formed from a catchment area 50~kpc in radius (this being the size
of the sphere we cut out of the previous simulation).

The growth rate is sufficiently slow that the halo responds
adiabatically, making the precise history of disc growth
irrelevant. Actual galaxy disks are thought to have been assembled
slowly (see, e.g., Sommer-Larsen \& Vedel \shortcite{SV98}).
At the end of the growth phase we allow the system to relax for a
period of $0.5$~Gyr during which the disc potential is kept constant.

\subsection{Disc blow-out}
\label{s:blow-out}

At this time the disc potential is instantaneously changed, in order
to mimic the rapid ejection of gas following a burst of star formation.
The total disc mass is reduced, and its surface density profile
adjusted to match published observational data on \DDO{154}.

We have found that a profile of the form
\begin{equation}
\Sigma(R) = \Sigma_0 e^{-(\sqrt{R^2+R_c^2}-R_c)/R_d}
\label{q:sjsl}
\end{equation}
with $\Sigma_0 = 5.9 \Msun/{\rm pc}^2$, $R_c=6.1$~kpc and
$R_d=1.0$~kpc provides a very good fit to the \HI{} data of Carignan
\& Freeman \shortcite{CF88} for a distance of 4~Mpc.
The corresponding total {\HI} mass within the radius probed by
Carignan \& Freeman is $2.63\times 10^8 \Msun$.

This only accounts for the gaseous component; for the observed stellar
disc remnant we adopt an exponential density profile, which gives a
reasonable fit to observations.

\begin{figure}
\epsfxsize=\hsize \epsfbox{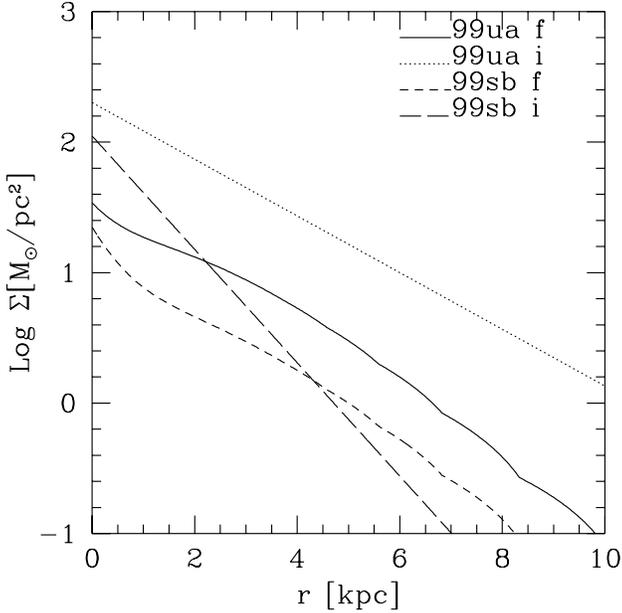}
\caption{Disc surface density profiles immediately before and after blow-out
(labels ``i'' and ``f'', respectively) for two representative simulation
runs. The profiles after blow-out include a stellar component in addition
to the gas.
}
\label{f:disc_sd}
\end{figure}

In figure~\ref{f:disc_sd} we show the run of surface density for the
combined gaseous and stellar disc components immediately before (``i''
curves) and after (``f'' curves) blow-out for two of our simulations.
The stellar component is only present in the post-blow-out discs, and
only manifests itself as a mass excess in the innermost 1~kpc.

\begin{figure}
\epsfxsize=\hsize \epsfbox{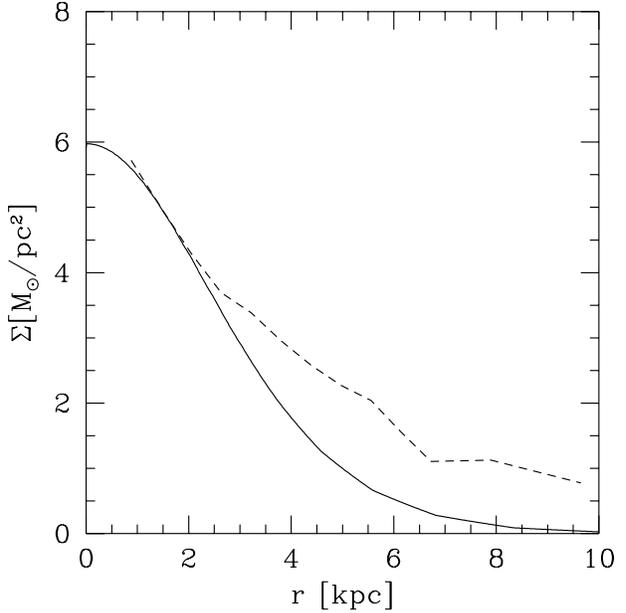}
\caption{Fit to the surface density profile of the \HI{} component
of \DDO{154} (solid curve), and critical surface density from
equation~(\ref{q:tsc}) assuming $\alpha=0.7$, $c_s=3.2$~km/s (dashed
curve). For these parameters the disc is stable everywhere. Larger
values of $\alpha$ and/or $c_s$ lead to larger maximum stable disc
masses. Typical observed values of $c_s$ in disk galaxies are between
6 and 10~km/s, allowing for a surface density 2 to~3 times higher than
shown in this figure even without invoking the uncertainty in $\alpha$.
}
\label{f:tsc}
\end{figure}

The disc of \DDO{154} may contain significant amounts of matter in
forms other than the detected \HI{} and stellar components.
A rough upper limit can be set by considering the stability of the
gaseous disc.
According to standard analyses of the problem, beginning with those of
Toomre \shortcite{T64} and of Goldreich \& Lynden-Bell
\shortcite{GL65}, the disc should be 
stable wherever its surface density lies below a critical value
\begin{equation}
\Sigma_c = \alpha { \kappa c_s \over 3.36 G}
\label{q:tsc}
\end{equation}
where $c_s$ is the velocity dispersion of the gas, $\kappa$ the
epicyclic frequency, and $\alpha$ a dimensionless constant near unity.
We follow Kennicutt \shortcite{K89} in deriving the epicyclic
frequency
\begin{equation}
\kappa = 1.41 {V\over R} \left(1 + {R\over V}{dV\over dR}\right)
\end{equation}
directly from the observed rotation curve.
Figure~\ref{f:tsc} compares the \HI{} surface density profile
(equation~[\ref{q:sjsl}], solid curve) with the critical surface density
$\Sigma_c$ (equation~[\ref{q:tsc}], dashed curve) for $\alpha=0.7$,
$c_s=3.2$~km/s.
These parameter values were chosen to make the surface density profile of
equation~(\ref{q:sjsl}) stable at all radii. They lie at the bottom end of
the plausible range of parameters: observed values of the velocity
dispersion~$c_s$ in disc galaxies typically range between 6 and 10~km/s
(unfortunately we could not find in the literature any measurement
specific to \DDO{154}), and the appropriate value for~$\alpha$ may also be
somewhat larger than~0.7.
The observational constraints therefore allow the disc of \DDO{154} to
be up to at least three times more massive than its observed \HI{}
content.

\subsection{Free parameters}

We allowed ourselves the following adjustments in the parameters of
the model:
\begin{enumerate}
\item varying the disc mass $M_{d,i}$ before blow-out;
\item varying the disc scale length $R_{d,i}$ before blow-out;
\item varying the ratio of total to {\HI} mass in the final disc.
\end{enumerate}

The scale length and mass of the disc before blow-out are not directly
observable quantities, and theoretical constraints on them are not
very tight, leaving us some leeway in choosing parameters for our models.
In general, smaller values of $R_{d,i}$ and larger masses will be more
effective in erasing the central cusp.
However, it would be remarkable if the initial disc structure were
totally unrelated to that of the observed stellar disc remnant.
Also, if we assume that the present gaseous disc is mostly made of
material that was present in the original disc before blow-out we
see that models with very small $R_{d,i}$ tend to be
ruled out by there being less angular momentum in the initial disc
than in the final one.
This is rather different from the assumption made by NEF, that
$R_{d,i}$ should be determined by the initial spin parameter
$\lambda\sim 0.05$ of the material. One should point out that the
expected distribution of spin parameter values is very broad and that
low surface brightness objects (like \DDO{154}) may very well be so
due to a higher degree of rotational support.
One way of reconciling these two views would be to assume a large
value of $M_{d,i}$ (which would allow $R_{d,i}$ to be decreased),
but very large gas masses would conflict with known constraints
on the abundance of baryons in the Universe.
We adopt an upper limit to the baryonic fraction of 0.3, consistent
with observational data on galaxy clusters. 
(With this choice, nucleosynthesis constraints can also be satisfied
for cosmic densities $\Omega \ga 0.3$ with $h=0.5$.)
None of our simulations requires
more than this fraction of the mass to be in baryonic form; most
require substantially less.

There is an arbitrariness in our choice of cosmological parameters:
the $\Omega=1$, $\Lambda=0$ standard CDM model is arguably not even the most
favoured by observations at the time of this writing. The times at
which we start and stop the disk growth were also chosen arbitrarily.
Fortunately it appears (NFW) that the structure of virialised haloes
is the same across a wide range of cosmologies; accordingly we need not
limit ourselves to parameter values (such as the total baryonic mass
discussed in the previous paragraph) compatible with $\Omega=1$
standard CDM.
The fact that our haloes were generated using a standard $\Omega=1$
CDM model should not be viewed as a severe limitation since the
choice of cosmology only affects the time of halo formation and the
relationship between concentration and mass, but not the internal
structure (NFW) or the internal dynamical evolution of the halo after
it has collapsed and virialised.

\section{Results}
\label{s:results}

\subsection{Initial halo profiles}

\begin{figure}
\epsfxsize=\hsize \epsfbox{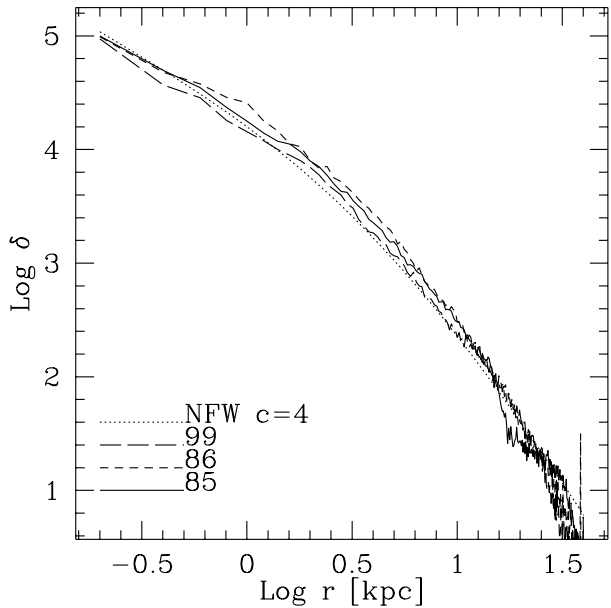}
\caption{Radial density profiles for the three selected haloes
(numbers 85, 86 and 99) at $t=3.0$~Gyr.
The dotted line shows the profile of equation~(\ref{q:nfw1}) for $c=4$
and $r_{200}=18$~kpc.
}
\label{f:rho3}
\end{figure}

\begin{figure}
\epsfxsize=\hsize \epsfbox{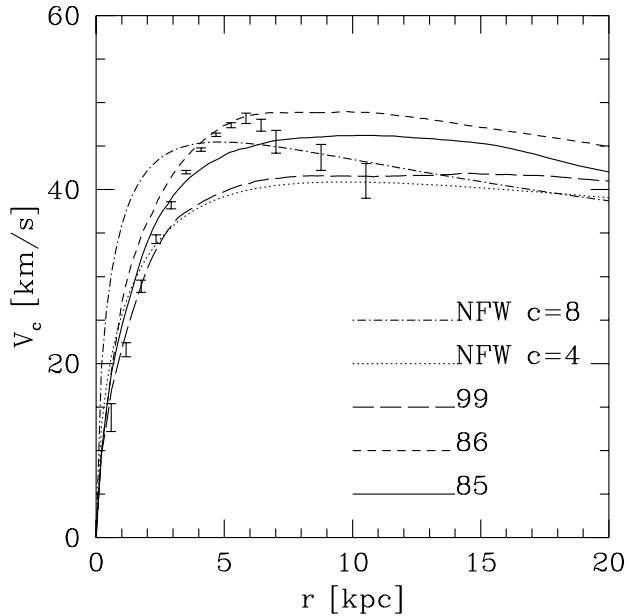}
\caption{Rotation curves for the three selected haloes at $t=3.0$~Gyr.
Dotted line as in figure~\ref{f:rho3}. Additionally, the dot-dash line shows 
the rotation curve for an NFW profile with concentration $c=8$.
The vertical error bars represent the observational data on \DDO{154}.
}
\label{f:vc3}
\end{figure}

Figure~\ref{f:rho3} shows the density profiles of our
three chosen haloes before disk growth, and compares them with a
fitting NFW profile (dotted line).
Figure~\ref{f:vc3} compares the corresponding rotation curves.
The virial mass of our haloes at this time is about $3\times 10^9
\Msun$. (They were selected to contain about $10^{10} \Msun$
at~$z=0$.)
NFW predict a concentration parameter $c\sim 7.8$ at $z=1.68$,
whereas the profiles in our simulations are fitted by $c\sim 4$.
(Optimum fits, in which the density in each bin is weighed by the
number of particles, actually yield a range of $c$ between 3.2 and~5
and $r_s$ between 3.6 and 5.5~kpc.)
The mass scale we are considering here lies outside the range studied
by NFW; it is possible that their formula gives poor results when
extrapolated.
We nonetheless examined four possible sources of error in our
simulations: collisional relaxation effects, softened gravity, neglect
of small-wavelength seed fluctuations, and finite time steps.

The rate of collisional relaxation is sensitive to the granularity
with which the mass is sampled. We repeated one of our runs with
2.35 times the original number of particles, and observed no
difference in the resulting profiles up to $t=4.2$~Gyr, at which point
we stopped the higher-resolution run.
We feel confident that two-body relaxation is not significantly
affecting our results.

We repeated another run with a softening length of 0.1 rather than
0.5~kpc, and again found the profile to be unchanged up to our final
simulation time (3.0~Gyr in this instance).
This, along with the observation that our characteristic radius $r_s
\equiv r_{200}/c$ exceeds the larger of the softening lengths by an
order of magnitude, leads us to rule out excessive softening as the
cause for our lower-than-expected value of~$c$.

There should be no need to add power down to the particle
Nyquist wavenumber.
The short-wavelength cutoff for our original $128^3$ grid in a $4
h^{-1}$~Mpc box corresponds to a characteristic mass of about
$1.7\times 10^7\Msun$, less than 1\% of the virial mass of our haloes.
Reproducing the NFW profile is expected to require resolving the
collapse of a progenitor with 1\% of the final virial mass;
our limiting wavelength is just small enough to achieve this.
Adding power on even smaller scales would force us to start the
simulations at a higher redshift, which in turn would require smaller
particle softenings and a larger number of particles to counter
collisional relaxation.

\begin{figure}
\epsfxsize=\hsize \epsfbox{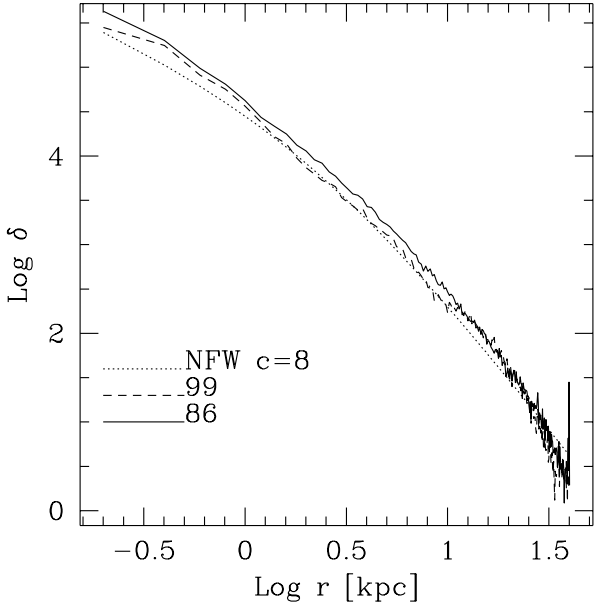}
\caption{Radial density profiles at $t=3.0$~Gyr for haloes 86 (solid)
and 99 (dashed), integrated with shorter time steps.
The dotted line shows the profile of equation~\ref{q:nfw1} for $c=8$
and $r_{200}=18$~kpc.
}
\label{f:rho3h}
\end{figure}

Although our changes to the time step selection criterion lead to a
more efficient distribution of individual particle time steps, we used
a value of the overall tolerance parameter at the high end of the
permissible range.
We therefore repeated one of our runs with a much smaller tolerance,
resulting in a sixteen-fold decrease in time step (from $4\times 10^6$
to $2.5\times 10^5$~years).
For the first Gyr, no difference was observed in the density
profiles with respect to the run with larger time steps; but starting
at a redshift $z\sim 4$ the results began to diverge. The less accurate
run evolved towards the shallow, $\rho \propto r^{-1}$ cusp of the NFW
profile while the new run maintained a central logarithmic slope much
closer to~$-2$.
A run with an intermediate choice of time step gave results in
agreement with the smaller-tolerance run, suggesting that results
may have converged (or else that they are now limited by a parameter
other than the time granularity of the simulation).
This seems to confirm the results of Fukushige \&
Makino~\shortcite{FM97}, and clearly deserves further investigation.
We plan to address this issue in more detail in a forthcoming paper.
A second run with initial conditions from one of our other haloes,
however, yielded a profile more similar to that of NFW with a
concentration $c\sim 8$. (This halo still displays a mass excess at
small radii, however. Figure~\ref{f:rho3h} shows the density profiles
for these two runs.) It appears therefore that at least some
haloes are reasonably (but not outstandingly) well fitted by NFW
profiles, and that for such 
haloes the concentration increases only moderately if smaller time steps
are used.
Moreover, we have no particular reason for favouring $t=3.0$~Gyr as
the time at which the disk began to form: we could equally have picked
an earlier time, when the concentration parameter $c$ was smaller
(but not {\em much} smaller, as that turns out to require unreasonably
high initial redshifts).
Or the true cosmological model could be one in which haloes in this
range of masses collapse later, and therefore have smaller $c$ values,
than in standard CDM. This is the case in particular for the more
favoured $\Lambda$-CDM models, in which $\Omega<1$ and the
cosmological constant $\Lambda$ is of the right magnitude to make the
model flat. Open CDM models may also be suitable in this perspective.

We therefore proceed with the analysis of our main series of runs,
having no serious grounds for rejecting $c\sim 4$ haloes as a valid
starting point for disk growth.
For completeness we also performed a few experiments with the more
concentrated haloes from the shorter time step integrations.
These experiments are computationally much more expensive than our
previous ones, and we could only perform a small number of them.
We shall briefly discuss the main differences in the results.

\subsection{Comparison with observational data}

We compare the rotation curves from our simulation runs with the
observations of Carignan \& Beaulieu~\shortcite{CB89}, as updated by
Carignan \& Purton \shortcite[as reported in BS]{CP97}.
For the purpose of estimating the \HI{} mass and the
characteristic length scale we place \DDO{154} at a distance of 4~Mpc,
consistent with photometric data \cite{CB89,HS95}.

The published rotation curve data are accompanied by error estimates;
we use these in computing a formal $\chi^2$ sum for each of our
simulated rotation curves.
We caution, however, against taking these $\chi^2$ values too
literally.
The measure relies on the published uncertainties in the rotation
curve, and is therefore dominated by the nominally better data from
the inner regions. It also assumes a Gaussian distribution of errors,
which may be unrealistic. Finally, it does not take into account the
scatter in the numerical models, which is expected to be larger in the
inner regions due to the smaller number of particles enclosed; this
suggests that one should perhaps give greater weight to the outer
part of the rotation curve than implied by the observational
uncertainties alone.

As an aid in assessing the overall quality of our fits, we note that
an isothermal sphere with a core radius of 3.0~kpc fits the data with
$\chi^2\sim 41$ if we first subtract the gaseous and stellar disc
components from the observed rotation curve (assuming that the error
bars are not affected).
While somewhat better than our best $\chi^2\sim 88$, these values
are significantly poorer than the $\chi^2\sim 13$ one would
expect in a truly outstanding fit with 13~degrees of freedom.
The quality of the fits to the rotation curve presented by BS is not
directly comparable to ours, since these authors effectively computed
an optimal mass distribution for their spheroid component, so that
the observed rotation curve is exactly reproduced by construction.

In our fits, we allow ourselves some freedom in rescaling the entire
rotation curve by a factor~$\alpha$, chosen optimally for each model.
The rescaling factor multiplies both radii and velocities. Masses
are implicitly multiplied by $\alpha^3$. The transformation leaves
dynamical times unchanged.
The justification for this rescaling is that we extract haloes from a
finite numerical simulation that samples a continuous spectrum of
halo properties. Our simulation may not have given us a halo of
exactly the right size, and the rescaling is meant to compensate for
this. It is naturally essential that the rescaled system remain
representative of the underlying halo population. For sufficiently
small adjustments a linear rescaling is adequate. We keep the
dynamical time constant so that the dynamical age, and hence the
internal structure, of each simulated halo is invariant under the
rescaling. This may mean that the rescaled halo corresponds to a
somewhat different redshift, but since we chose the redshift of disk
formation somewhat arbitrarily in the first place this has no
adverse impact on the quality of our results. In the ideal case
of a singular isothermal sphere, the virial radius at a given
redshift is linearly proportional to the circular velocity, just as in
our adopted rescaling. (See for example equation~2 of Mo, Mao \& White
\shortcite{MMW98}.)

Models for which $\alpha$ differs considerably from unity
should be regarded with caution regardless of their nominal $\chi^2$
value: the rescaling applies also to the \HI{} and stellar disc
components, which were initially chosen to match observational data with
$\alpha=1$ and no longer agree with these observational data
after rescaling by an $\alpha$ very different from unity.

We can also justify rescaling the radial coordinate independently of
the circular velocity: both the distance to \DDO{154} and the Hubble
constant carry uncertainties of at least 10--20\%. We therefore also
present the 
result of rescaling radii by a factor $\beta$ and velocities by a
factor $\gamma \ne \beta$; the best $\chi^2$ values then become
comparable to that for the isothermal sphere fit.
\begin{table*}
\begin{minipage}{100mm}
\caption{Characteristics of the runs}
\label{t:runs}
\begin{tabular}{lrlrrllrllr}
Run	& $M_{d,1}$	& $R_{d,1}$	& $M_{d,2}$	& 
$M_{H,2}$	& $R_{d,2}$	& $\alpha$	& $\chi^2_{\nu=13}$	&
$\beta$		& $\gamma$	& $\chi^2_{\nu=12}$	\\
	& $[M_{\HI}]$	& [kpc]		& $[M_{\HI}]$	&
$[M_{\HI}]$	& [kpc]		&		&		&
		&		&		\\
(1)	& (2)	& (3)	& (4)	& (5)	& (6)	& (7)	& (8)	& (9)
& (10)	& (11)	\\
\hline	
85\ddag	&	&	&	&	&	& 1.06	& 241	& 1.59	& 1.18	& 143	\\
86\ddag	&	&	&	&	&	& 0.96	& 283	& 1.35	& 1.05	& 167	\\
99\ddag	&	&	&	&	&	& 1.19	& 179	& 1.45	& 1.25	& 101	\\
99sa	& 3.5	& 1	& 1.2	& 1	& 0.5	& 1.13	& 152	& 0.94  & 1.06	& 76	\\
99sb	& 2.7	& 1	& 1.2	& 1	& 0.5	& 1.06	& 89	& 1.02	& 1.05	& 87	\\
99sc	& 1.8	& 1	& 1.2	& 1	& 0.5	& 1.01	& 136	& 1.06	& 1.02	& 118	\\
99ta	& 6.4	& 2	& 3.2	& 3	& 0.5	& 0.89	& 326	& 1.22	& 0.95	& 44	\\
99ua	& 9.6	& 2	& 3.2	& 3	& 0.5	& 0.98	& 165	& 1.19	& 1.02	& 46	\\
85a	& 9.6	& 2	& 3.2	& 3	& 0.5	& 1.00	& 297	& 1.35	& 1.07	& 60	\\
85b	& 6.4	& 2	& 3.2	& 3	& 0.5	& 0.91	& 476	& 1.35	& 0.99	& 61	\\
85e	& 12.8	& 2	& 3.2	& 3	& 0.5	& 1.00	& 230	& 1.27	& 1.06	& 55	\\
86a	& 9.6	& 2	& 3.2	& 3	& 0.5	& 0.97	& 266	& 1.28	& 1.03	& 61	\\
86b	& 6.4	& 2	& 3.2	& 3	& 0.5	& 0.89	& 385	& 1.27	& 0.96	& 54	\\
86c	& 3.5	& 2	& 1.2	& 1	& 0.5	& 1.00	& 188	& 1.25	& 1.06	& 113	\\
86d	& 1.8	& 2	& 1.2	& 1	& 0.5	& 0.94	& 349	& 1.45	& 1.04	& 119	\\
99Ca\dag& 9.6	& 0.5	& 1.2	& 1	& 0.5	& 1.42	& 563	& 1.96	& 1.41	& 196	\\
86Ea\dag& 9.6	& 0.5	& 1.2	& 1	& 0.5	& 1.27	& 338	& 3.33	& 1.71	& 169	\\
\hline
\end{tabular}
\end{minipage}
\end{table*}

Table~\ref{t:runs} summarizes the properties of our main series of
runs. Column~1 contains the name of each run; the leading
digits identify the halo realisation used as a starting point.
Runs marked with a dagger (\dag) correspond to a higher initial
concentration $c\sim8$ and smaller time steps.
Runs marked with a double dagger (\ddag) represent the dark matter halos
before disk growth and removal.
Column~2 lists the maximum mass of the exponential disc that has
been grown (in units of $M_{\HI}\equiv2.63\times 10^8 \Msun$, the
estimated mass 
of the observed \HI{} disc in \DDO{154}), and column~3 the
scale length of this exponential disc (in~kpc).
The total mass of the remnant disc after blow-out appears in
column~4, and the mass in the \HI-like component in column~5.
Column~6 shows the exponential scale length of the stellar
component of the remnant disc; the mass of this component is
given by subtracting column~5 from column~4, and amounts to $5\times
10^7 \Msun$ for all the runs presented here. We treat all disc
components as having infinite extent.
The $\chi^2$ value in column~8 is for $\nu=13$ degrees of freedom,
corresponding to the 14 data points of the observational rotation
curve, together with the fact that we allowed ourselves a rescaling of
the rotation curve by a factor~$\alpha$ (shown in column~7).
Columns 9, 10 and~11 correspond to independent rescaling of the radii
by a factor $\beta$ (column~9) and velocities by $\gamma$
(column 10), yielding $\chi^2_{\nu=12}$ in column~11.

\begin{figure}
\epsfxsize=\hsize \epsfbox{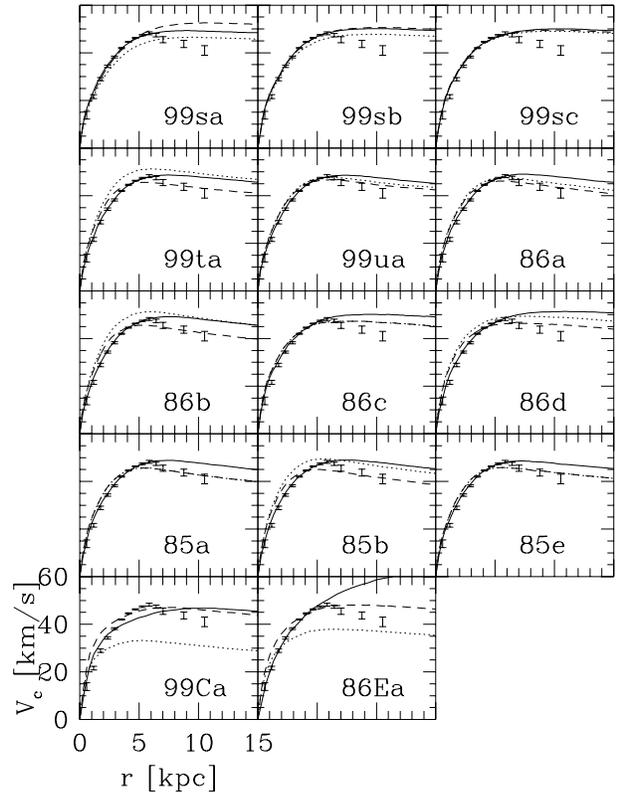}
\caption{Fits to the rotation curve for the runs listed in
table~\ref{t:runs}. Error bars are centred on the observational data
points. Each tile is labelled with the name of the corresponding run.
The dotted curve shows the unscaled rotation curve. The dashed curve
corresponds to our one-parameter rescaling, the solid curve to our
two-parameter rescaling.
}
\label{f:vc14}
\end{figure}

The $\chi^2$ values alone cannot capture all the relevant information
about the quality of the fits.
We therefore also present plots of the rotation curves for all our
runs in figure~\ref{f:vc14}.
Each tile in this figure is labelled with the name of the
corresponding run from column~1 of table~\ref{t:runs}.
The error bars are centred on the observational data points.
The dotted curve represents the raw (unscaled) rotation curve from the
simulation run; the dashed curve corresponds to rescaling both $r$ and
$v$ by $\alpha$ (table~\ref{t:runs}, column~7), the solid curve to
rescaling $r$ by $\beta$ (column~9) and $v$ by $\gamma$ (column~10
of table~\ref{t:runs}).

The quality of the fits appears to be relatively insensitive to the
amount of mass loss (compare runs 99sa, 99sb and~99sc; 85a, 85b
and~85e; 86a and~86b; 86c and~86d), at least within the range (33\% to
75\% of the gas mass before blow-out) spanned by our main series of 
simulations.

It appears difficult to reconcile the observed decline in the circular
velocity at large radii without producing some mass excess at small
radii.
This mass excess can however be contained within reasonable limits.
Not unexpectedly, increasing the mass in the remnant gas disk improves
the fit to the outer parts of the rotation curve; however, nearly all
our models (and certainly all those which provide decent fits to the
rising parts of the rotation curve) fall somewhat short of reproducing
the observed decline.
Given the size of the observational error bars and the idealisations
in our model (notably the assumption of instantaneous mass
redistribution), however, it is not
clear to us that this should be a major source of concern.
There is some scatter between our individual halo realizations
(compare runs 99ua, 85a, 86a), suggesting that the observations are at
least compatible with the overall scenario embodied in our
simulations, if not precisely reproduced by the particular halo
realizations we tried. 

We have also performed a few experiments with more centrally
concentrated ($c\sim 8$) initial haloes.
(These would be more appropriate models to use if the true cosmology is
$\Omega=1$, $\Lambda=0$; on the other hand, low-$\Omega$ models
can produce less concentrated haloes more in
line with our $c\sim 4$ runs.)
The clear outcome of these runs (99Ca, 86Ea) is that even if we
exaggerate the strength of halo pinching by making the disk more
massive and more centrally concentrated during the growth phase, we
are barely able to produce a central core of the required size, and
the overall fit to the observed rotation curve is extremely poor.
We must therefore conclude that our scenario only explains the
observed rotation curve of \DDO{154} if the effective concentration of
the dark matter halo was closer to $c=4$ than to $c=8$.

Burkert~\shortcite{B95} has pointed out that the core structure of the
dark matter halo in~\DDO{154} appears to be shared by other galaxies
in its class, and questioned the ability of the mass ejection
hypothesis to produce the fine-tuning that he claims is necessary to
turn NFW's one-parameter family of cuspy $N$-body halo profiles into
the different one-parameter family of observed profiles.
In this regard we would like to point out that we obtain reasonable
fits to the rising part of the rotation curve for a rather wide range
of ejected mass fractions; the real challenge is rather to fit the
{\em declining} part of \DDO{154}'s rotation curve.
Of the rotation curves in Burkert's~\shortcite{B95} sample, only
\DDO{154}'s extends sufficiently far outwards to provide such a
stringent test.
While his argument about fine-tuning is undoubtedly interesting, we do
not find it entirely compelling given the current paucity of
observational data.

\section{Conclusions}
\label{s:conclusions}

Our numerical experiments confirm that it is not very easy to give a
satisfactory account of the observed rotation curve of \DDO{154} within
the framework of a CDM cosmogony.
It is, however, possible to obtain a reasonable fit if one assumes
\begin{enumerate}
\item that sudden mass ejection from the central star-forming regions
of the galaxy was responsible for erasing the cusp predicted by CDM
models of collisionless gravitational collapse;
\item that the disc of \DDO{154} is about three times more massive
than implied by its 21~cm \HI{} emission;
\item that the effective halo concentration was low.
\end{enumerate}

We are grateful to Julio Navarro for having made his numerical code
available, to Sasha Kashlinsky for early encouragement to look
further into this problem, and to Per Rex Christensen for kindly
proofreading the manuscript.
We also thank the referee for his comments which helped us improve the
presentation of this paper.
This work was supported by Danmarks Grundforskningsfond through its grant
for the establishment of the Theoretical Astrophysics Center.

\end{document}